\documentclass[referee]{aa}
\usepackage{graphicx}
\usepackage{times} 
\newcommand{\kms} {~km~s$^{-1}$}
\newcommand{\vlsr}  {$v_{\rm LSR}$}
\newcommand{\J}[2]  {\mbox{$J$=#1$\to$#2}}
\renewcommand{\sun}   {\odot}
\newcommand{\msun}{{\,\rm M}_{\odot}}
\begin{document}




\title{L1157: Interaction of the molecular outflow with the Class~0
environment}

\author{M.T.~Beltr\'an\inst{1,2} \and F.~Gueth\inst{3} \and
S.~Guilloteau \inst{3} \and A.~Dutrey \inst{4}}

\institute{
 Harvard-Smithsonian Center for Astrophysics, 60 Garden Street,
 Cambridge, MA 02138, USA 
 \and now at Osservatorio Astrofisico di Arcetri,
 Largo E.\ Fermi 5, I-50125 Firenze, Italy 
 \and Institut de Radio Astronomie Millim\'etrique,
 300 Rue de la Piscine, F-38406 Saint Martin d'H\`eres, France
 \and Laboratoire d'Astrophysique de l'Observatoire de
 Grenoble, BP 53, F-38041 Grenoble, France
 }

\date{Received , Accepted }
\authorrunning{Beltr\'an et al.}
\titlerunning{The environment of L1157}


\abstract{We present high angular resolution interferometric
observations [\thanks{Based on observations carried out with the IRAM
Plateau de Bure Interferometer.  IRAM is supported by INSU/CNRS
(France), MPG (Germany) and IGN (Spain).}] of the dust continuum at
2.7 and 1.3~mm, and of the HC$_3$N~(\J{12}{11}) and
C$^{18}$O~(\J{2}{1}) emission around L1157-mm, a Class~0 object that
drives a spectacular molecular outflow. The millimeter dust emission
is clearly resolved into two components, a flattened compact source of
$\sim$450$\times$250~AU at 1.3~mm, and mass $\sim$0.1~$M_\sun$, plus
an extended envelope of $\sim$3000~AU at 1.3~mm, and mass
$\sim$1.1~$M_\sun$.  The millimeter spectral index varies throughout
the region, with the lower value found toward the compact protostar,
possibly indicating grain growth in the denser regions. A strong
interaction between the molecular outflow and the close protostellar
environment is taking place and affects the structure of the innermost
parts of the envelope. This is shown by the spatial coincidence
between the molecular outflow and the dust (1.3~mm continuum) and
HC$_3$N emission: both tracers show structures associated to the edges
of the outflow lobes. Basically, the global picture sketched for the
Class~0 object L1157-mm by Gueth et al.\ (1997) is supported. We find
possible evidence of infall, but we do not detect any velocity
gradient indicative of a rotating circumstellar disk.}

\maketitle


\section{Introduction}

Class~0 low-mass young stellar objects are deeply embedded in
circumstellar dust and gas material, and are found associated with
very energetic molecular bipolar outflows. Theory outlines a scenario
where a central object is surrounded by an infalling envelope that
contains most of the mass (e.g.\ Larson 1969; Adams, Lada, \& Shu
1987). This infalling material is accreted onto the central protostar
funneled through a circumstellar disk that grows as the system evolves
(e.g.\ Cassen \& Moosman 1981; Terebey, Shu, \& Cassen 1984; Shu,
Adams, \& Lizano 1987). On the other hand, a powerful outflow is
accelerated and collimated via a magnetohydrodynamic mechanism from
the surface of the accretion disk (Pudritz \& Norman 1983; Heyvaerts
\& Norman 1989; Shu et al.\ 1994; K\"onigl \& Pudritz 2000).
Infalling, outflowing and rotational motions are, thus, simultaneously
taking place in extremely young protostar environments, making the
morphology and kinematics of such regions very complex. Observations
of the millimeter thermal dust continuum and high-density molecular
tracers can, in principle, reveal the phenomena taking place in the
innermost parts of dense cores (e.g.\ Saito et al.\ 1996; Ohashi et
al.\ 1997, 1999; Lai \& Crutcher 2000; Hogerheijde 2001). However, the
strong interaction between the outflow and the surrounding dense
material, which has been actually observed near the driving source in
a few cases (e.g.\ L1551-IRS5: Fuller et al.\ 1995, Ladd et al.\ 1995;
L1157: Gueth et al.\ 1997; L1527: Ohashi et al.\ 1997, Hogerheijde et
al.\ 1998, Motte \& Andr\'e 2001), complicates this portrait.  In
order to get an accurate picture of the complex environment of Class~0
objects and to interpret the gas kinematics in these sources, high
angular resolution interferometric observations of different tracers
are definitely required.

L1157-mm is a Class~0 object located at 440~pc with a $L_{\rm
bol}\simeq11\,L_\sun$. It is associated with IRAS~20386+6751 and
drives a spectacular outflow. The L1157 outflow has been studied in
details through many molecular lines, such as CO (Ume\-mo\-to et al.\
1992; Gueth, Guilloteau, \& Bachiller 1996; Bachiller \& P\'erez
Guti\'errez 1997; Hirano \& Taniguchi 2001), SiO (Zhang et al.\ 1995,
2000; Gueth, Guilloteau, \& Bachiller 1998; Bachiller et al.\ 2001),
H$_2$ (Hodapp 1994; Davis \& Eisl\"offel 1995), NH$_3$ (Bachiller,
Mart\'{\i}n-Pintado, \& Fuente 1993; Tafalla \& Bachiller 1995), or
CH$_3$OH (Bachiller et al.\ 1995, 2001; Avery \& Chiao 1996). Many
other lines have been detected (Bachiller \& P\'erez Guti\'errez 1997;
Bachiller et al.\ 2001), making L1157 the prototype of chemically
active outflows.  Regarding the protostar itself, dust continuum
observations have been carried out at 2.7~mm (Gueth et al.\ 1996,
1997), 1.3~mm (Shirley et al.\ 2000; Motte \& Andr\'e 2001; Chini et
al.\ 2001; Gueth et al.\ 2003), 850~$\mu$m (Shirley et al.\ 2000;
Chini et al.\ 2001), and 450~$\mu$m (Chini et al.\ 2001).

In a first attempt to understand the morphology of this young object,
Gueth et al.\ (1997) carried out $^{13}$CO~(\J{1}{0}),
C$^{18}$O~(\J{1}{0}), and 2.7~mm continuum IRAM Plateau de Bure
interferometric observations of the region around L1157-mm, with a
$\sim$2$\farcs$5 angular resolution. The continuum emission clearly
shows two components: a marginally resolved compact, flattened core,
perpendicular to the outflow direction, and a low-level extended
emission that seems to delineate the heated edges of the cavity
excavated by the CO outflow. The $^{13}$CO emission also originates
from the limb-brightened edges of the outflow, whereas the C$^{18}$O
emission is more directly associated with the compact continuum
source, and shows marginal evidence of rotation.  In addition,
redshifted self-absorption is present in the $^{13}$CO spectrum, which
suggests infall motions. However, in order to confirm the picture
described by Gueth et al.\ (1997), higher angular resolution and
higher frequency continuum observations, together with different,
higher density, tracers were necessary. In this work we present these
new interferometric observations of the dust continuum at 2.7 and
1.3~mm, and of the HC$_3$N~(\J{12}{11}) and C$^{18}$O~(\J{2}{1})
emission around L1157-mm.  These new maps reveal important details of
the structure of the central source, and of the interaction between
the outflow and the high-density core.

\section{Observations}

\subsection{Interferometric data}

Observations were carried out with the IRAM Plateau de Bure
Interferometer (Guilloteau et al.\ 1992) between November 1995 and
April 1996.  Five different configurations of the four-antennas array
were used. The longest baseline was 288~m (B2) and the shortest 24~m
(B2, D). The phase center was set to the position
$\alpha($J$2000)=20^{\rm h}39^{\rm m}06\fs19$,
$\delta($J$2000)=68\degr02'15\farcs9$, which is the L1157-mm position
reported by Gueth et al.\ (1997). The source was observed
simultaneously at 109.2~GHz (HC$_3$N~\J{12}{11}) in USB and 219.6~GHz
(C$^{18}$O~\J{2}{1}) in LSB band. The spectra were analyzed using a
correlator with one band of 10~MHz centered on the
HC$_3$N~(\J{12}{11}) line, one band of 20~MHz centered on the
C$^{18}$O~(\J{2}{1}) line, and two bands of 160~MHz for the 2.7 and
1.3~mm continuum. The spectral resolution was 0.11~km\,s$^{-1}$ in
both narrow bands. The bandpass of the receivers was calibrated by
observations of the strong quasars 3C273, 3C454.3, 0923+392, or
1823+568.  Amplitude and phase calibrations were achieved by
monitoring 1823+568, whose flux density was determined relative to
3C273 and 3C454.3.

However, the emission of 1823+568 turned out to be linearly polarized,
and therefore presented a sinusoidal dependence of the amplitude with
time i.e.\ with parallactic angle. Special considerations had thus to
be taken during the calibration process.  Since the polarization of
the two receivers (2.7 and 1.3~mm) is orthogonal, the amplitude curve
A($t$) of a linearly polarized source has an opposite behavior on both
receivers. This effect is only no\-ti\-ce\-able in tracks of several
hours. To fix this polarization problem, we first ran a procedure to
fit the calibrator amplitude curve with a sinusoidal function of the
parallactic angle $\Psi$, for each scan. The functions were
proportional to 1+C1$\times\cos^2(\Psi+\Phi)$, for the 2.7~mm
receiver, and to 1+C2$\times\sin^2(\Psi+\Phi)$ for the 1.3~mm
receiver, where C1 and C2 give an estimate of the percentage of
polarized flux, and $\Phi$ gives an idea of the polarization
angle. Then, we calibrated the amplitude curve against one of the
well-known calibrators also used in the observations (e.g.\ 3C273), in
order to obtain the real value of the flux for each scan. Thus,
instead of a constant value for the flux of the amplitude calibrator,
we obtained a function of the parallactic angle $\Psi$, and this
function is the one we used in the calibration process.

We estimate that the uncertainties on the absolute flux density
calibration are $\sim$20\% at 2.7~mm, and $\sim$40\% at 1.3~mm, due to
the difficulties encountered to calibrate the amplitude. Typical rms
phase noise was better than $10\degr$ at 2.7~mm, and better than
$20\degr$ at 1.3~mm.  The data were calibrated and analyzed with the
GILDAS software package de\-ve\-lo\-ped at IRAM and Observatoire de
Grenoble.  The synthesized CLEANed beam for maps made using natural
weighting was $2\farcs24\times2\farcs04$ at P.A.~$= 33\degr$ at
2.7~mm, and $1\farcs24\times1\farcs14$ at P.A.~$= 71\degr$ at 1.3~mm.
Unless explicitly mentioned, we subtracted the continuum from the line
emission. This was performed directly in the $uv$ plane in order to
avoid non linearity effects in the deconvolution, and thus any
amplification of errors induced in this process.

\subsection{Bolometer data}

Gueth et al.\ (2003) presented 1.3~mm bolometer measurements obtained
with the MPIfR 19-channel bolometer array (MAMBO) on the IRAM 30-m
telescope. We used this map to extract short-spacings information and
thus complement the interferometric data set. The interferometer and
30-m data at 1.3~mm were merged in the $uv$ plane using the procedure
available in the GILDAS package. The relative weights of the
interferometer and single-dish data were adjusted so that the weight
density of the single-dish data matches that of the shortest
interferometric baselines. A multiplicative scaling factor of 0.8 was
applied to the single-dish data amplitudes, in order to compensate for
the fact that the single-dish observations were carried out at a
central wavelength of $\sim$1.25~mm, whereas the interferometric
observations were indeed done at 1.4~mm. Hereafter, we will refer to
the combined interferometer plus single-dish 1.3~mm continuum data
simply as 1.3~mm continuum data. The synthesized (natural weight)
CLEANed beam for the resulting map, $1\farcs24\times1\farcs22$ at $\rm
P.A.=27\degr$, is very similar to the beam without the bolometer data.

\begin{figure*}[!ht]
  \centerline{\resizebox{16.0cm}{!}{\includegraphics[angle=270]{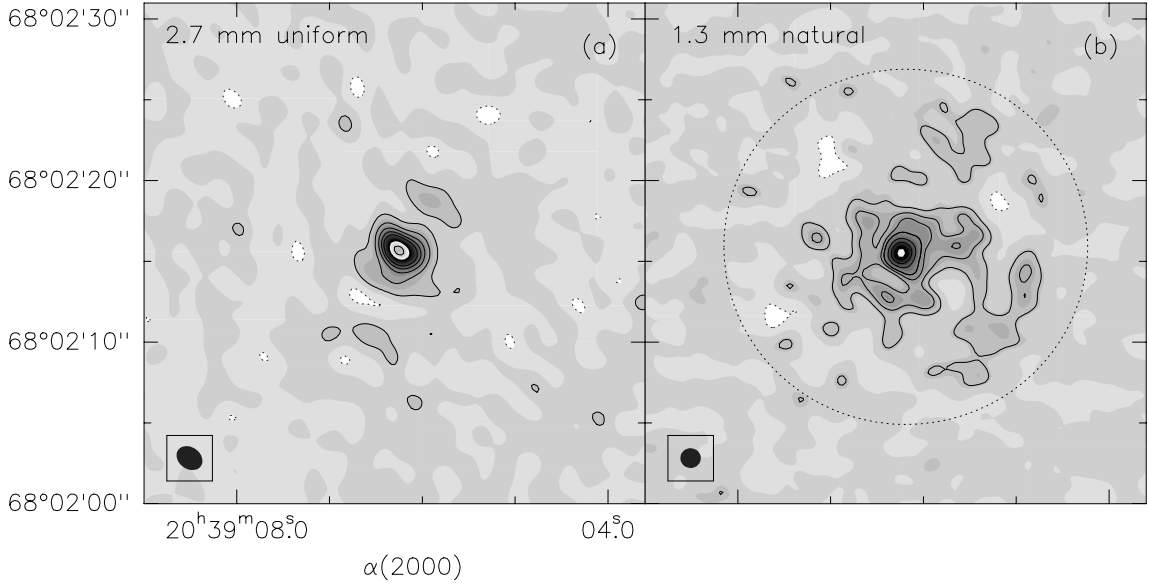}}}
  \caption{ \label{fig1} $(a)$ 2.7~mm continuum emission
  uniform-weight map. The synthesized beam is
  $1\farcs70\times1\farcs33$ at $\rm P.A.=38\degr$, and is drawn in
  the bottom left corner. The rms noise of the map is
  0.33~mJy\,beam$^{-1}$. The contour levels are $-$1, 1, 3, 5, 7, 9,
  13, and 17~mJy\,beam$^{-1}$. ($b$) 1.3~mm continuum natural-weight
  map after adding the single-dish data. The synthesized beam is
  $1\farcs24\times1\farcs22$ at $\rm P.A.=27\degr$, and is drawn in
  the bottom left corner. The rms noise of the map is
  1.4~mJy\,beam$^{-1}$. The contour levels are $-$5, 5, 10, 15, 20,
  30, 40 and 50~mJy\,beam$^{-1}$. The circle represents the Plateau
  de Bure primary beam (50\% attenuation level).}
\end{figure*}

\section{Continuum emission}

\begin{figure*}[!ht]
  \centerline{\resizebox{15.0cm}{!}{\includegraphics[angle=270]{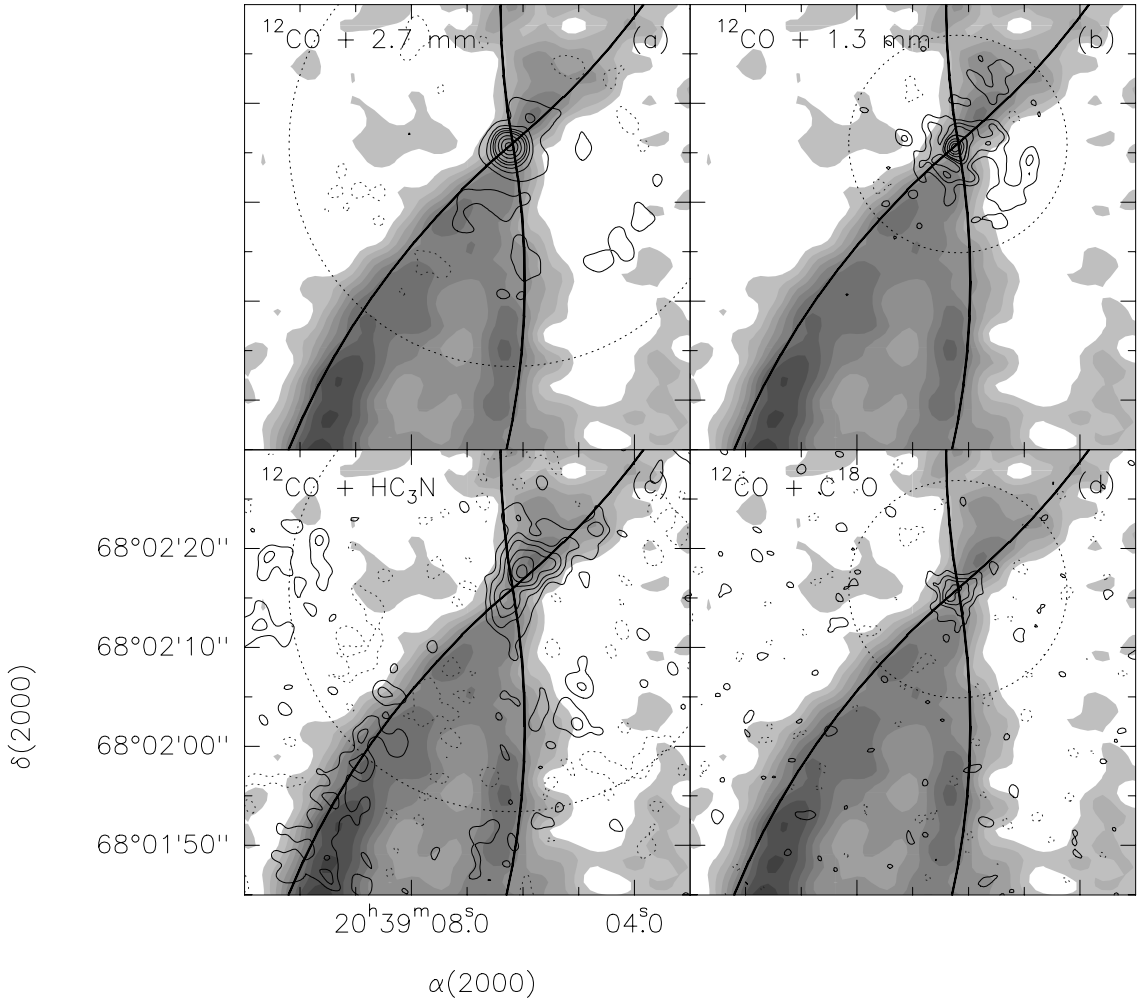}}}
  \caption{\label{fig2} Overlay of the $^{12}$CO~(\J{1}{0}) integrated
  emission ({\it greyscale}; adapted from Gueth et al.\ 1996) and
  $(a)$ the 2.7~mm continuum natural-weight map toward L1157 ({\it
  contours}). The rms noise of the map is 0.30~mJy\,beam$^{-1}$. The
  contour levels are $-$1, 1, 3, 5, 7, 9, 13, 17, and
  21~mJy\,beam$^{-1}$;
  $(b)$ the 1.3~mm continuum natural-weight map after adding the
  single-dish map data ({\it contours}). The rms noise of the map is
  1.4~mJy\,beam$^{-1}$.  The contour levels are $-$5, 5, 10, 15, 20,
  30, 40 and 50~mJy\,beam$^{-1}$; 
  $(c)$ the HC$_3$N~(\J{12}{11}) integrated emission ({\it
  contours}). The rms noise of the map is 4.8~mJy\,beam$^{-1}$\kms\
  (0.11~K\kms). The contours are $-$12, 12, 24, 36, 48, 60 and
  72~mJy\,beam$^{-1}$\kms; 
  and $(d)$ the C$^{18}$O~(\J{2}{1}) integrated emission 
  ({\it contours}). The rms noise of the map is 16~mJy\,beam$^{-1}$\kms\ 
  (0.28~K\kms). The contours are $-$70, 70, 140, 210, and 
  280~mJy\,beam$^{-1}$\kms. 
  The HC$_3$N and the C$^{18}$O emission have been integrated over the
  outflow velocity range 1.8 to 3.6\kms. The black lines outline the
  edges of the $^{12}$CO outflow, for reference in other figures.  The
  circles represent the Plateau de Bure primary beam (50\% attenuation
  level).}
\end{figure*}

The millimeter continuum emission towards the core of L1157 was mapped
at 2.7 and 1.3~mm. Figure~\ref{fig1} shows the uniform-weight map of
the 2.7~mm emission, and the natural-weight map of the 1.3~mm
emission. Figure~\ref{fig2} shows the superposition of the integrated
$^{12}$CO~(\J{1}{0}) emission (adapted from Gueth et al.\ 1996), which
is tracing the molecular outflow, with the 2.7~mm and 1.3~mm
natural-weight maps (as well as with the HC$_3$N and C$^{18}$O
integrated emission).  The continuum dust emission is resolved at both
wavelengths and clearly shows two components, a compact source plus an
extended envelope. In Table~\ref{tmm}, we summarize the observed
properties of these two components. The position found for the compact
source L1157-mm at both wavelengths is $\alpha($J$2000)=20^{\rm
h}39^{\rm m}06\fs24$, $\delta($J$2000)=68\degr02'15\farcs6$. Position
accuracy is about $0\farcs2$. This is in agreement with the position
found by Gueth et al.\ (1996) through previous 115~GHz
observations. The position obtained by Gueth et al.\ (1997) is
slightly different, but this was traced back to to an incorrect value
for the coordinates of the phase calibrator 2013+370. 

\begin{table*}[!hbt]
\caption{Properties of the compact and extended continuum emission \label{tmm}}
\begin{tabular}{|l|c|c|c|c|c|c|c|c|c|}
\hline & \multicolumn{3}{c|}{Integrated fluxes} &
\multicolumn{3}{c|}{Source size$^c$} & \multicolumn{2}{c|}{Spectral indexes}& Masses\\
Component & 2.7~mm$^\mathrm{a}$ & 2.7~mm$^\mathrm{b}$ & 1.3~mm$^\mathrm{b}$ & & & & & &  \\
&(mJy) & (mJy) & (mJy) & ($\arcsec$) & (AU) & P.A. & $\alpha$ & $\beta=\alpha-2$ & ($\msun$)\\
 \hline
 COMPACT & 35 & $25^\mathrm{d}$ & $78^\mathrm{e}$
 & $1.05\times0.58^\mathrm{f}$ &$\sim$450$\times$250 &12$\degr$ & $\sim$2.1 & 0.1 & $\sim$0.12\\
 \hline
EXTENDED & $\sim$60 & $\sim$31 & $\sim$680 &$\sim$8$\times$7& $\sim$3500$\times$3100 & $\sim$155$\degr$ & 4--5 & 2--3 & $\sim$1.08 \\
 \hline
 TOTAL$^\mathrm{g}$ & 95 & 56 & 760 & & & & & & $\sim$1.2 \\
\hline
\end{tabular}\\
Peak coordinates of the compact core: $\alpha$(J2000) =
20$^{\rm h}$39$^{\rm m}$06.24$^{\rm s}$ $\delta$(J2000)
 = 68$\degr$02$'$15.6$''$. Position accuracy is $0\farcs2$. \\
(a) Integrated flux density measured by Gueth et al.\ (1997).\\
(b) Integrated flux density measured in this paper, corrected for
primary beam response. \\
(c) Measured at 1.3~mm. \\
(d) Integrated flux density measured by fitting an elliptical
Gaussian directly to the
visibility data for baselines $\geq 80$~m. \\
(e) Integrated flux density measured by fitting an elliptical
Gaussian directly to the
visibility data for baselines $\geq 40$~m. \\
(f) Size obtained by fitting an elliptical Gaussian directly to the
visibility data. The
uncertainty in the size values is $\sim$0$\farcs$04. \\
(g) Compact + extended components. \\
\end{table*}

\begin{figure*}[!th]
  \centerline{\resizebox{16.0cm}{!}{\includegraphics[angle=270]{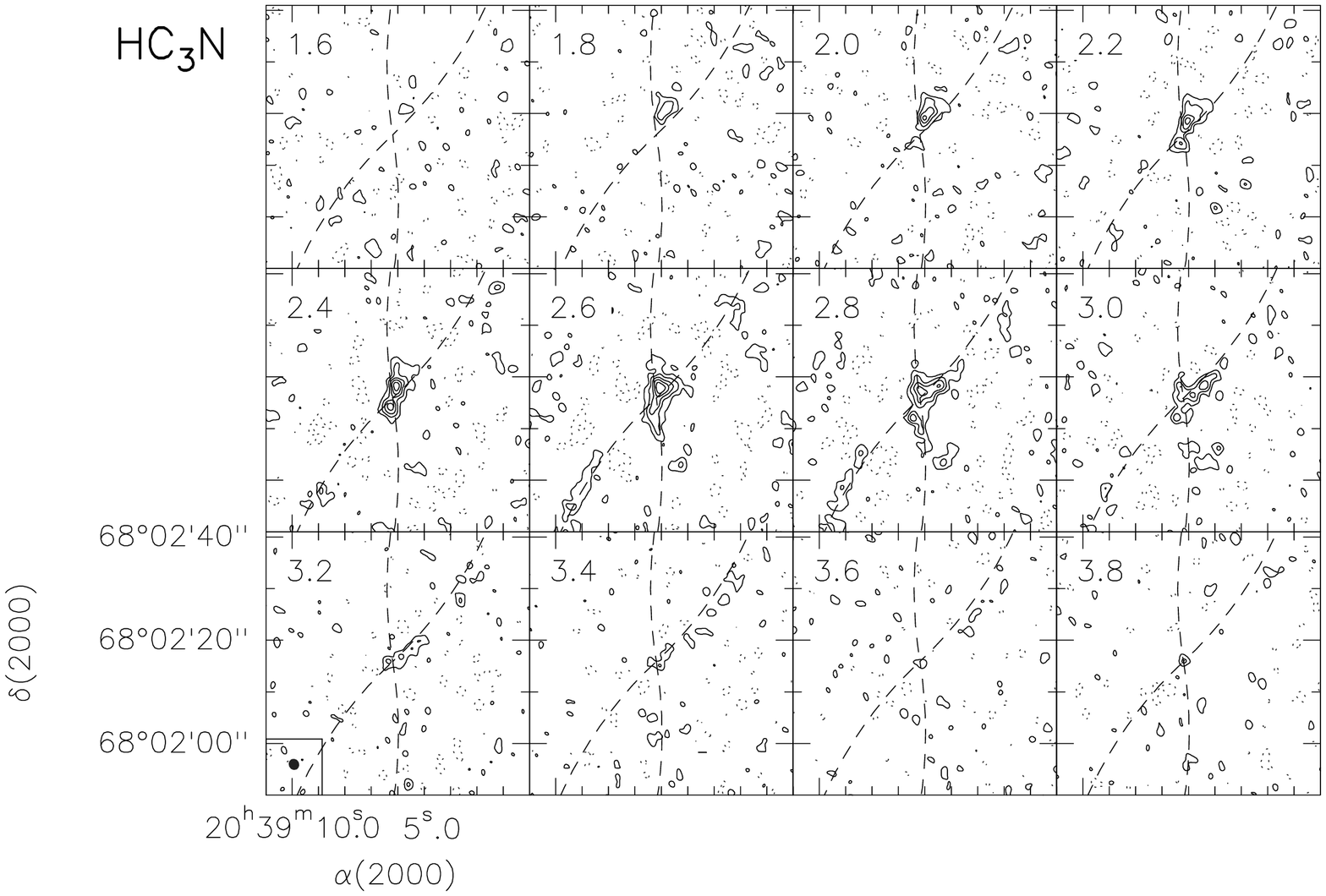}}}
  \caption{ \label{fig3} Velocity channel maps of the
  HC$_3$N~(\J{12}{11}) emission. The \vlsr\ of L1157-mm is
  2.6\kms. The central outflow velocity of each velocity interval is
  indicated at the upper left corner of each panel. The 1-$\sigma$
  noise in 1 channel is 7.5~mJy\,beam$^{-1}$.  Contour levels are
  $-$15, 15 to 75 by step of 15~mJy\,beam$^{-1}$. The conversion
  factor from Jy\,beam$^{-1}$ to K is 22.5. The synthesized beam is
  $2\farcs24\times2\farcs04$ at $\rm P.A.=33\degr$, and is drawn in
  the bottom left corner of the bottom left panel. The dash lines
  outline the edges of the outflow (see Fig.~\ref{fig2}).}
\end{figure*}

\begin{figure*}[!t]
  \centerline{\resizebox{16.0cm}{!}{\includegraphics[angle=270]{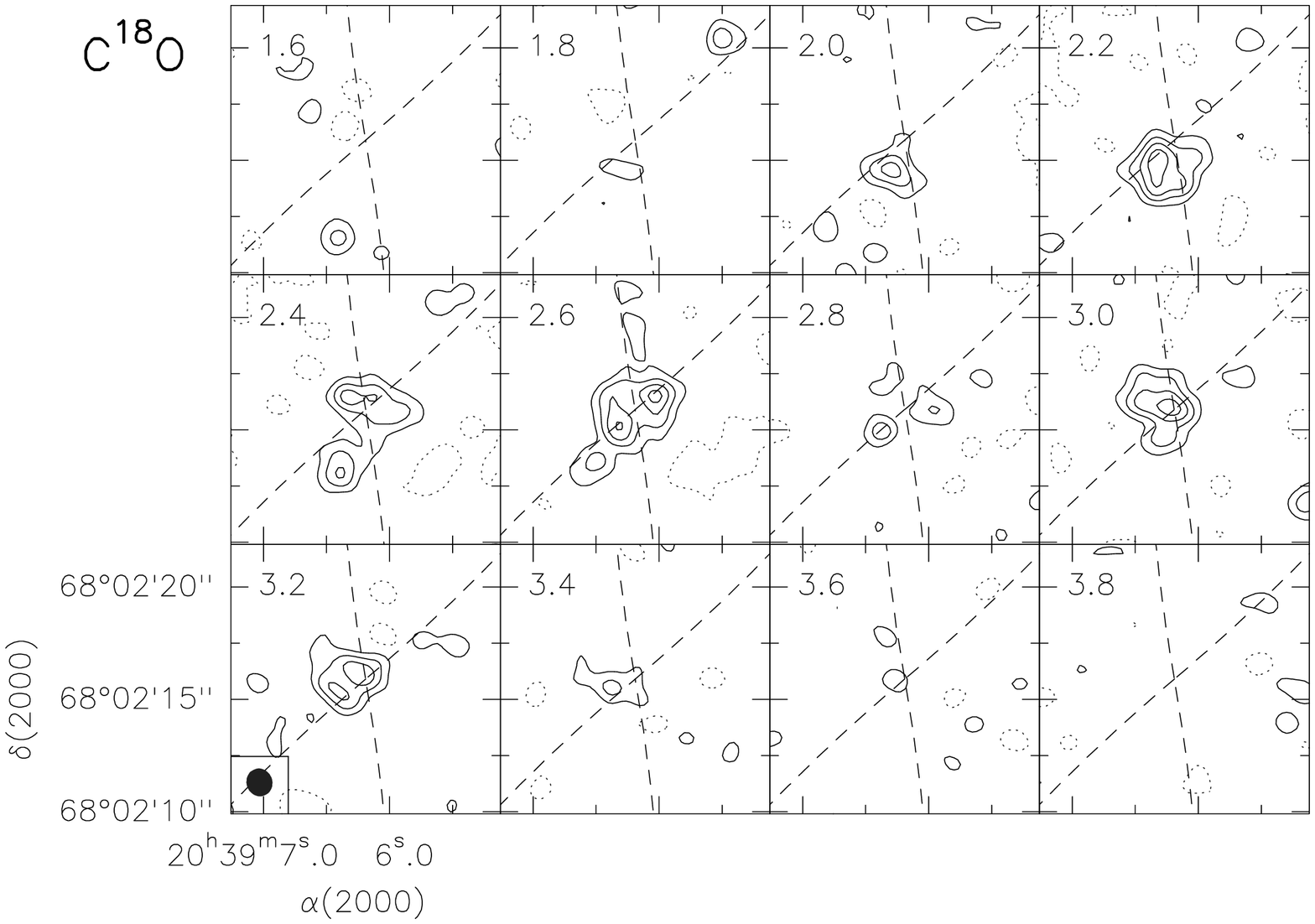}}}
  \caption{ \label{fig4} Velocity channel maps of the
  C$^{18}$O~(\J{2}{1}) emission. The \vlsr\ of L1157-mm is
  2.6\kms. The central outflow velocity of each velocity interval is
  indicated at the upper left corner of each panel. The 1-$\sigma$
  noise in 1 channel is 30~mJy\,beam$^{-1}$.  Contour levels are
  $-$50, 50 to 200 by step of 50~mJy\,beam$^{-1}$. The conversion
  factor from Jy\,beam$^{-1}$ to K is 17.9. The synthesized beam is
  $1\farcs24\times1\farcs14$ at $\rm P.A.=71\degr$, and is drawn in
  the bottom left corner of the bottom left panel. The dash lines
  outline the edges of the outflow (see Fig.~\ref{fig2}).}
\end{figure*}

\subsection{2.7~mm continuum}
\label{2.7mm}

The 2.7~mm uniform-weight map (Fig.~\ref{fig1}) marginally resolves
the compact component: the central source is slightly elongated
perpendicular to the outflow direction, but this elongation is however
dominated by the beam. The deconvolved size we found after fitting an
elliptical Gaussian to the uniform-weight map is
$\sim$1$\farcs3\times1\farcs2$ at P.A.\ $\simeq50\degr$, which is
different from the position angle of the uniform-weight synthesized
beam (P.A.\ $=38\degr$). Fitting an elliptical Gaussian directly to
the visibility data, for baselines longer than 80~m, we found a total
flux of 25~mJy and a size of
$1\farcs2\pm0\farcs04\times0\farcs9\pm0\farcs05$ at P.A.\
$=50\degr$. This size, which corresponds to a linear size of
$\sim$500$\times$400~AU at the distance of the source, is consistent
with the previous determination from Gueth et al.\ (1997). However,
the flux density value that we have found is slightly lower than the
value of 35~mJy measured by these authors.

Extended emission at a low 1 to 3~mJy\,beam$^{-1}$ level is also
visible along the outflow axis at P.A.\ of $\sim$155$\degr$. This
emission has a flux density of $\sim$31~mJy integrated over an area of
$\sim$200~arcsec$^2$. It is less extended and weaker than the emission
detected by Gueth et al.\ (1997) with a similar synthesized
beam. Adding the compact component, the total continuum flux at 2.7~mm
is $\sim$56~mJy. This is lower than the flux of $\sim$90~mJy measured
by Gueth et al.\ (1997). These discrepancies reflect the difficulty of
absolute flux calibration in the mm domain, as well as that of
low-level extended structures deconvolution -- which may crucially
depend on the actual $uv$ coverage.

\subsection{1.3~mm continuum}
\label{1.3mm}

At 1.3~mm, the compact component has a peak intensity of
60~mJy\,beam$^{-1}$, and is mar\-gi\-nal\-ly resolved, with a size of
$\sim$1$\farcs1\times0\farcs9$. By fitting an elliptical Gaussian to
the visibility data for baselines longer than 40~m (which is
consistent with using 80~m at 2.7~mm in order to select the same
physical scales), we obtained a total flux of $\sim$78~mJy, and a size
of $1\farcs05\pm0\farcs04\times0\farcs58\pm0\farcs04$
($\sim$450$\times$250~AU) at P.A.\ $=12\degr$. Note that both the size
and orientation of the compact source are slightly different at 2.7
and 1.3~mm, which is probably related to the source not having a
perfect Gaussian shape structure. In particular, at 1.3~mm the compact
component is slightly elongated {\em along} the outflow cavity.

Figures~\ref{fig1} and~\ref{fig2} show that the centrally peaked
compact source is surrounded by a structure extended roughly along the
flow axis. The flux density is 5 to 15~mJy\,beam$^{-1}$ in this
component, well above the rms noise, $\sigma\simeq
1.4$~mJy\,beam$^{-1}$ (estimated over an empty area of the map). We
measured a total flux density of $\sim$760~mJy (corrected for primary
beam attenuation). This is consistent with the flux of 630~mJy
measured at 1.1~mm by Motte \& Andr\'e (2001) with MAMBO on the IRAM
30-m telescope. The circumstellar dust emission around L1157-mm is
clearly dominated by the extended component. This envelope does not
have spherical symmetry, but it is spread over a region of $\sim$8$''$
($\sim$3500~AU) along the outflow direction. 

The 1.3~mm map also shows an emission extended in the direction
perpendicular to the outflow, on a $\sim$20$''$ scale. This structure
is obviously much too large to be a circumstellar disk. It might be
due to a flattened envelope remnant of the molecular cloud in which
the protostar was formed. This extended material has also been
detected, but at a much larger spatial scale, through 850~$\mu$m and
1.3~mm single-dish observations (Shirley et al.\ 2000; Chini et al.\
2001; Gueth et al.\ 2003).

\section{Molecular line emission}

\subsection{HC$_3$N maps}

The HC$_3$N (\J{12}{11}) transition has a high critical density
($8\times10^5$~cm$^{-3}$, Chung, Osamu, \& Masaki 1991), which makes
it an even better tracer of high density regions than e.g.\
CS~(\J{1}{0}) and CS~(\J{2}{1}). Figure~\ref{fig3} shows the velocity
channel maps for the HC$_3$N~(\J{12}{11}) emission toward the core of
L1157, around the systemic velocity channel ($\simeq2.6$\kms). The gas
emission around L1157-mm is not compact, but extended and elongated
along the outflow edges. The outflow is blueshifted toward the south
and redshifted toward the north and is almost in the plane of the sky
(Gueth et al.\ 1996). As a consequence, there is blueshifted and
redshifted HC$_3$N emission in both lobes. The integrated HC$_3$N
emission (Fig.~\ref{fig2}) has a flux density
$\sim$670~mJy\,km\,s$^{-1}$.

At a distance of $\sim$30$''$ from the central source, the southern
lobe reveals an elongated filament at velocity 2.6 and
2.8~km\,s$^{-1}$ (Fig.~\ref{fig3}). It is outlining very clearly the
eastern flank of the outflow CO lobe. This emission is already
significantly affected by the primary beam attenuation. Interestingly,
this feature coincides with a SiO shock (S3 shock from Gueth et al.\
1998; Zhang et al.\ 2000), traced also by H$_2$ (Davis \& Eisl\"offel
1995), NH$_3$ (Tafalla \& Bachiller 1995), H$_2$CO, CS, CH$_3$OH, and
SO (Bachiller et al.\ 2001). Gueth et al.\ (1996, 1998) and Zhang et
al.\ (2000) describe a precessing episodic jet scenario that can
explain the different orientations and velocities of the cavities of
the outflow. According to this scenario, the jet has precessed toward
the SE--NW direction, and the region discussed here would thus
correspond to an event of enhanced ejection, which is impacting the
flank of the existing CO cavity. The resulting shock explains the
increased abundances of some species at that location, including
HC$_3$N (Bachiller \& P\'erez-Guti\'errez 1997). The velocity of the
HC$_3$N gas is significantly lower than that of the CO emission at the
same location (e.g.\ Gueth et al.\ 1996). This suggests that high
velocity material has a lower density, and would thus not be seen in
HC$_3$N, although we cannot rule out an opacity effect.
No HC$_3$N emission has been detected inside the cavity, but an
extended emission could possibly have been filtered out by the
interferometer.

\subsection{C$^{18}$O maps}

Figure~\ref{fig4} presents the velocity channel maps for the
C$^{18}$O~(\J{2}{1}) emission. The integrated emission
(Fig.~\ref{fig2}) appears more compact than the HC$_3$N emission: it
is dominated by the central compact component, and is thus associated
to the innermost part of the protostellar envelope.  The total
integrated intensity is $\sim$3~Jy\,km\,s$^{-1}$. The C$^{18}$O
emission is also tracing gas associated to the outflow, as can be seen
at the systemic velocity channel map (\vlsr$ =2.6$~km\,s$^{-1}$, see
Fig.~\ref{fig4}). A striking feature visible in Fig.~\ref{fig4} is
that the C$^{18}$O emission at 2.8~km\,s$^{-1}$ is weaker than in the
previous and following velocity channels. This effect is clearly seen
as a narrow deep self-absorption in the spectrum taken at the position
of L1157-mm (see below, section~\ref{self} and Fig.~\ref{fig6}).

\section{Analysis}

\subsection{Spectral index map}

\begin{figure}[!th]
  \resizebox{\columnwidth}{!}{\includegraphics[angle=270]{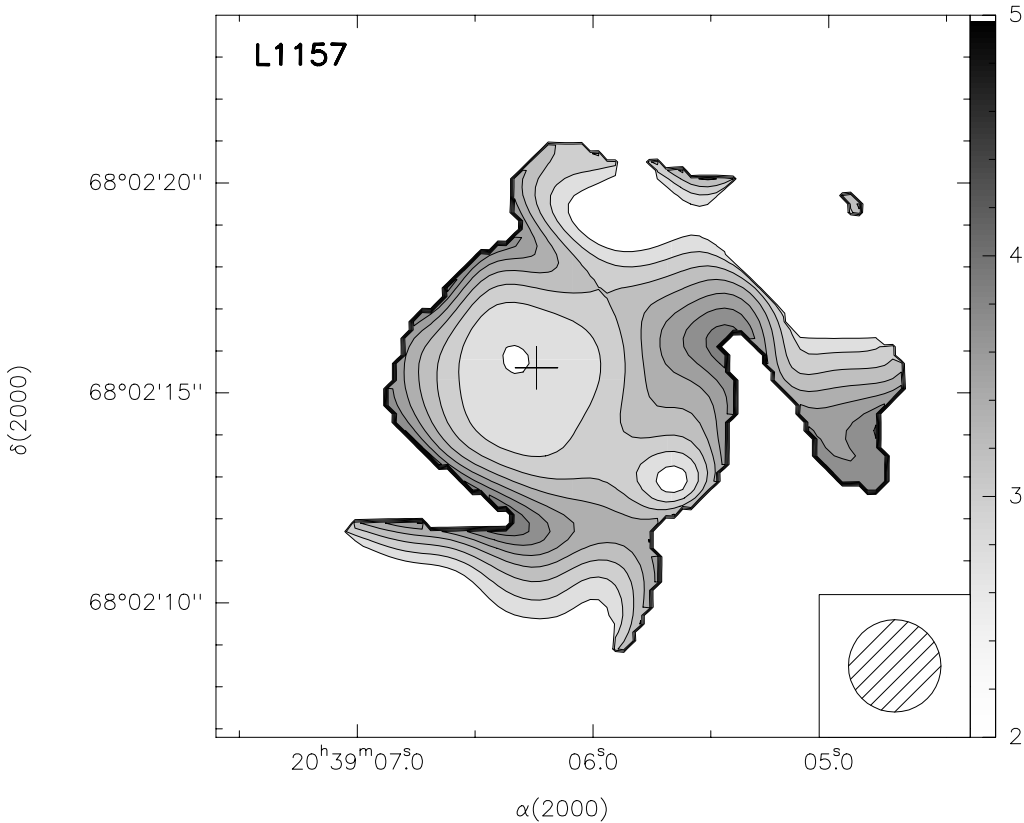}}
  \caption{ \label{fig5} Spectral index map from the 2.7 and 1.3~mm
  continuum images, restored to the same resolution of
  $2\farcs2$. Contour levels are plotted from 2 to 5 by intervals of
  0.5. Greyscale is linear in the range of 2--5, and shows the regions
  with higher spectral index darker. The cross marks the position of
  L1157-mm. }
\end{figure}

Figure~\ref{fig5} presents a map of the spectral index $\alpha$, where
$S_\nu\propto\nu^\alpha$, computed using the 2.7 and 1.3~mm continuum
interferometric images restored with the same resolution, $2\farcs2$
(or $\sim$900~AU). The spectral index varies from a value $\sim$2.1,
at the position of the compact source to a value of $\sim$4--5 at
larger distance\footnote{Gueth et al.\ (2003) have analyzed the
spectral index around L1157-mm from bolometric observations at 1.2 and
0.85\,mm. They obtained an intermediate value of $\alpha \simeq 3.5$
in the central 20$''$ beam which includes all our interferometric
map.}. Instrumental effects such as missing flux and/or a wrong
flux scale would affect the absolute level of the spectral index map,
but can hardly introduce a spatial variation. Line contamination can
be excluded, since it does not affect the interferometer data and was
found to be absent in the bolometer measurement (Gueth et al.\ 2003).
The observed variation accross the L\,1157-mm region may thus be
indicative of physical differences between the properties of the
material toward the embedded source and those of the more extended
region. This effect was already observed in other young embedded
sources (Visser et al.\ 1998; Johnstone \& Bally 1999; Smith et al.\
2000; Beltr\'an et al.\ 2002).

A low value of $\alpha$, as observed at the central position, can be
due to optically thick dust emission. In such a case, the dust
temperature $T_d$ is equal to the brightness temperature $T_B$ (in the
Rayleigh-Jeans approximation), which can be derived from the observed
flux density:
\begin{equation}
S_\nu=\frac{2\,k\,\nu^2}{c^2}\,T_B\,\Omega,
\end{equation}
where $\Omega$ is the solid angle subtended by the source. An upper
limit on $T_d$ can be computed when assuming that the whole flux of
the compact core (78 mJy, see Table 1) is coming from the optically
thick contribution. Assuming that the size of the optically thick
emitting region is around 150~AU, we derive a value of $T_d<10$~K for
the dust temperature (note that the larger the optically thick core,
the lower $T_d$). This value is very low and suggests that a
significant fraction of the compact core has an optically thin
emission. 

Assuming that the dust emission at mm wavelengths is optically thin
and in the Rayleigh-Jeans regime, $\alpha$ is related to the power-law
index $\beta$ of the dust emissivity $\kappa_\nu\propto\nu^\beta$,
through $\alpha=\beta+2$. For the more extended material, $\beta$
would thus be larger than $2$, the typical value for interstellar dust
grains (see e.g.\ Draine \& Lee 1984), while the value of $\beta$ for
the compact source would be $\sim$0.1. Interestingly, this decrease of
$\beta$ could be due to larger grains (e.g.\ Mannings \& Emerson
1994), hence suggesting grain growth in the denser inner
regions. Grain shape evolution or chemical evolution are also possible
(see, e.g.\ Ossenkopf \& Henning 1994; Pollack et al.\ 1994).

\subsection{Envelope mass -- from continuum emission}

Assuming optically thin dust thermal emission, the mass of the
emitting material can be derived from the 1.3~mm continuum flux
density $S_\nu$ by:
\begin{equation}
M=\frac{S_\nu\,D^2}{\kappa_\nu\,T}\,\frac{c^2}{2\,k\,\nu^2}
\end{equation}
Using a dust tem\-pe\-ra\-tu\-re of 40~K (following Gueth et al.\
1997), and a dust mass opacity of 0.01~cm$^2$\,g$^{-1}$, as
re\-commen\-ded by Henning, Michel, \& Stognienko (1995) for a
gas-to-dust ratio of 100, we obtain a mass of the central compact
source of $\sim$0.12~$M_\sun$, while the mass of the envelope is
$\sim$1.08~$M_\sun$. Thus, the total (compact + extended)
circumstellar mass is $\sim$1.2~$M_\sun$. This is a factor 3 below the
value of $\sim$3$M_\sun$ derived by Gueth et al.\ (1997) at
2.7\,mm. The absolute flux density of the 1.3~mm data may be
significantly underestimated. These calculations are also dependent on
the dust opacity law used, the mass being lower in case of higher dust
opacity. In any case, the large ratio between the mass of the extended
component and that of the compact one indicates that there is an
important reservoir of material that can be incorporated into the
central protostar. This is in agreement with the conceptual definition
of a Class~0 object.

\subsection{Envelope mass -- from gas emission}

The gas mass of the protostellar condensation can be better estimated
from the C$^{18}$O data, since the HC$_3$N emission is partly
associated to the outflow. Following the derivation of Scoville et
al.\ (1986), and assuming optically thin emission, the C$^{18}$O beam
averaged column density is given by
\begin{equation}
\label{density} 
\bar N=1.15\,\,10^{13}\, \frac{(T_{\rm ex}+0.88)}{e^{-15.81/T_{\rm
ex}}}\int{T_B\,dv} ~~{\rm cm}^{-2}
\end{equation}
where $T_{\rm ex}$ is the excitation temperature, and $\int{T_B dv}$
is the integrated brightness temperature of the C$^{18}$O~(\J{2}{1})
emission. Assuming $T_{\rm ex}=40$~K, a C$^{18}$O abundance relative
to molecular hydrogen of $1.4\times10^{-7}$, and a mean molecular
weight of $2.6\,m_H$, we obtain a mass of
$\sim$9$\times10^{-3}~M_\sun$ in the inner $\sim$2$''\times2''$ area
(roughly corresponding to the central source). It should be mentioned
that the excitation temperature value adopted does not affect
significantly the estimated molecular mass; the mass is
$\sim$7$\times10^{-3}~M_\sun$ for $T_{\rm ex}=20$~K, and
$\sim$11$\times10^{-3}~M_\sun$ for $T_{\rm ex}=50$~K. Our derived
value is certainly a lower limit since the C$^{18}$O is affected by
self-absorption (see $\S$~\ref{self}), and hence partly optically
thick. Indeed, Gueth et al.\ (1997) derived a mass of
$4.7\times10^{-2}~M_\sun$ from C$^{18}$O~(\J{1}{0}). Comparing the
molecular mass, a few $10^{-2}~M_\sun$, to that derived from the
con\-ti\-nuum data, $M\simeq0.12~M_\sun$, it suggests that CO in
L1157-mm could be depleted by a factor around $\sim$3--10, although
this result has to be taken with caution in case the gas is optically
thick and/or clumpy. Using the same assumptions as above, the mass of
the whole C$^{18}$O envelope ($\sim$8$''\times8''$ area) is
$\sim$4$\times10^{-2}~M_\sun$, again significantly below the mass
derived from the continuum emission (1.08~$M_\sun$). However, the
C$^{18}$O emission is less extended that the continuum emission, and
the comparison between the two masses may therefore not be
relevant. It is also possible that we miss some C$^{18}$O~(\J{2}{1})
emission because of the lack of short spacings in our measurement.

\section{Discussion}
\label{discus}

\begin{figure*}[!t]
  \includegraphics[angle=270,width=11cm]{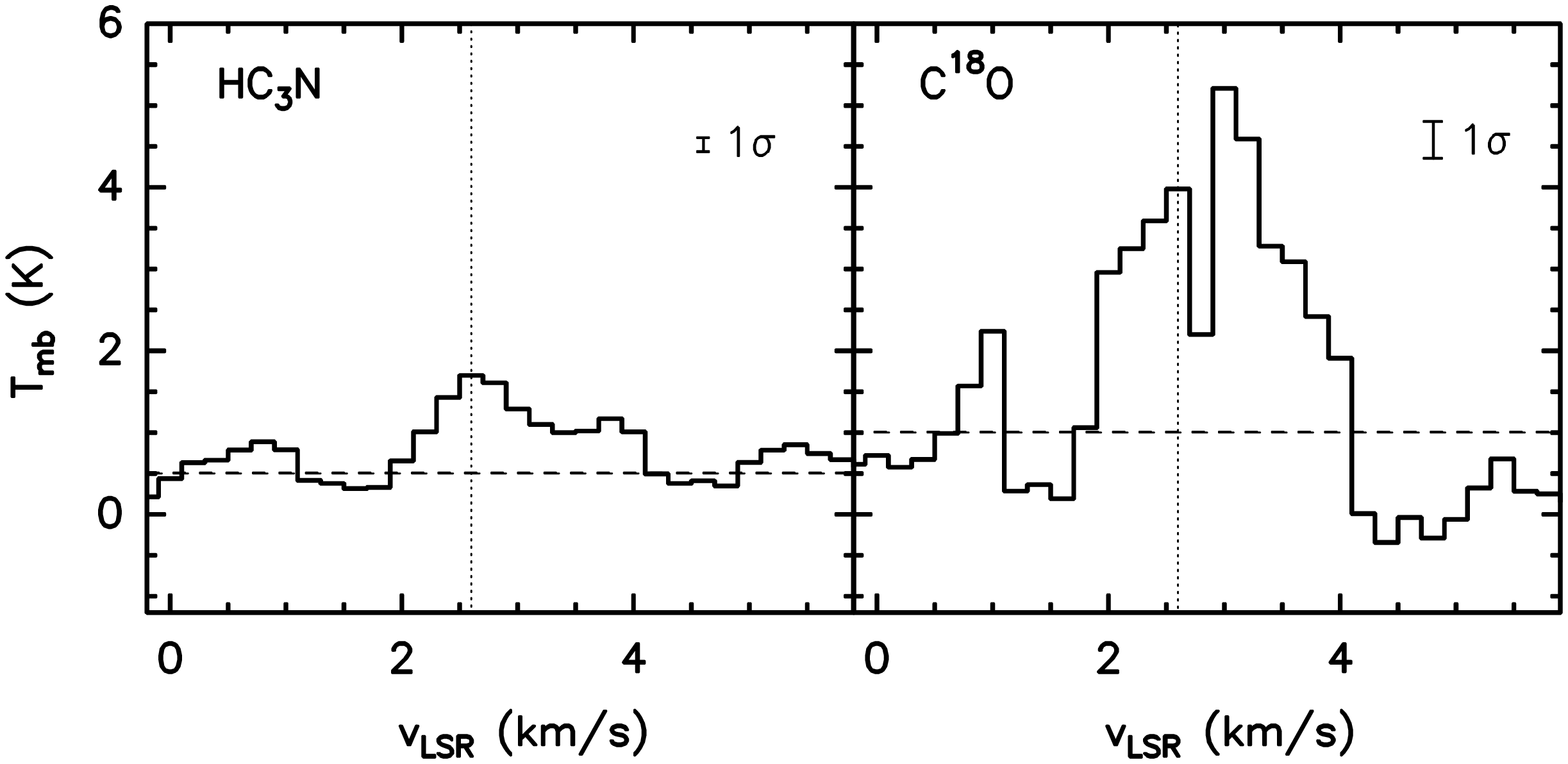} \hspace{1cm}
  {\caption{ \label{fig6} HC$_3$N~(\J{12}{11}) ({\it
  left}) and C$^{18}$O~(\J{2}{1}) ({\it right}) spectra obtained at
  the position of L1157-mm. The continuum emission has not been
  subtracted (the continuum level is indicated by the dashed
  horizontal line). The dashed vertical line indicates the \vlsr\ of
  2.6\kms. The 1-$\sigma$ noise in 1 channel is 0.17~K for HC$_3$N,
  and 0.45~K for C$^{18}$O. The conversion factor of Jy\,beam$^{-1}$
  to K is 22.5 for HC$_3$N, and 17.9 for C$^{18}$O. }}
\end{figure*}

\subsection{Interaction between the outflow and the envelope}

Both our continuum and lines observations reveal structures or
filaments that seem to be associated with the edges of the L\,1157
outflow.

The 1.3~mm continuum map (Fig.~\ref{fig2}) reveals a low-level
extended emission whose conical shape matches that of the outflow
lobes. Gueth et al.\ (1997) reported a similar feature from 2.7~mm
observations: the dust emission was observed and modelled as a
``cross-like'' structure coincident with the outflow edges. Note
however that these observations indicated opening angles slightly
smaller than those shown by our maps. The morphological association
between the outflow edges and dense protostellar material in L\,1157
is even better shown by the the HC$_3$N observations. The channel maps
presented in Fig.~\ref{fig3} (see the 2.6 to 3.2~\kms\ channels)
reveal extensions that outline quite accurately the beginning of the
outflow conical lobes. The southeastern arm of this ``cross-like''
pattern is missing, as in the 2.7~mm continuum from Gueth et al.\
(1997), which suggests different temperature and/or density than in
the other arms.  The C$^{18}$O emission, while much more compact than
the HC$_3$N emission, exhibits marginal indications of a similar
morphology.

Following Gueth et al.\ (1997), we suggest that these morphological
coincidences indicate that a strong interaction between the outflow
and the protostellar envelope is taking place in L\,1157: the outflow
sweeps up envelope material, thereby creating heated/compressed
regions at its edges, which, observationaly, mimic a ``cross-like''
structure (see Gueth et al.\ 1997, their Fig.~10). Circumstellar dust
and gas material around young stellar objects outlining the edges of
the molecular outflow have also been mapped in other sources (e.g.\
B5-IRS1: Langer, Velusamy, \& Xie 1996; L1527: Ohashi et al.\ 1997,
Hogerheijde et al.\ 1998, Motte \& Andr\'e 2001; L1551-IRS5: Fuller et
al.\ 1995, Ladd et al.\ 1995).

Interestingly, the HC$_3$N emission also revels two peaks located on
both sides of the central protostar.  They are clearly visible in the
integrated emission map (Fig.~\ref{fig2}) as well as in the individual
channel maps (Fig.~\ref{fig3}). Their positions coincide with the
apparent base of the outflow conical cavities as traced by the
$^{12}$CO (see Fig.~\ref{fig2}), suggesting that these emission peaks
could be produced when the dense gas is shock-heated and compressed by
the flow. Hence, we conclude that, at the resolution of our
observations, the structure of the dust+gas protostellar envelope is
strongly affected by the presence of the outflow.

\subsection{Kinematics of the Gas}

\subsubsection{C$^{18}$O Self-absorption}
\label{self}

\begin{figure}[!t]
  \resizebox{\columnwidth}{!}{\includegraphics[angle=270]{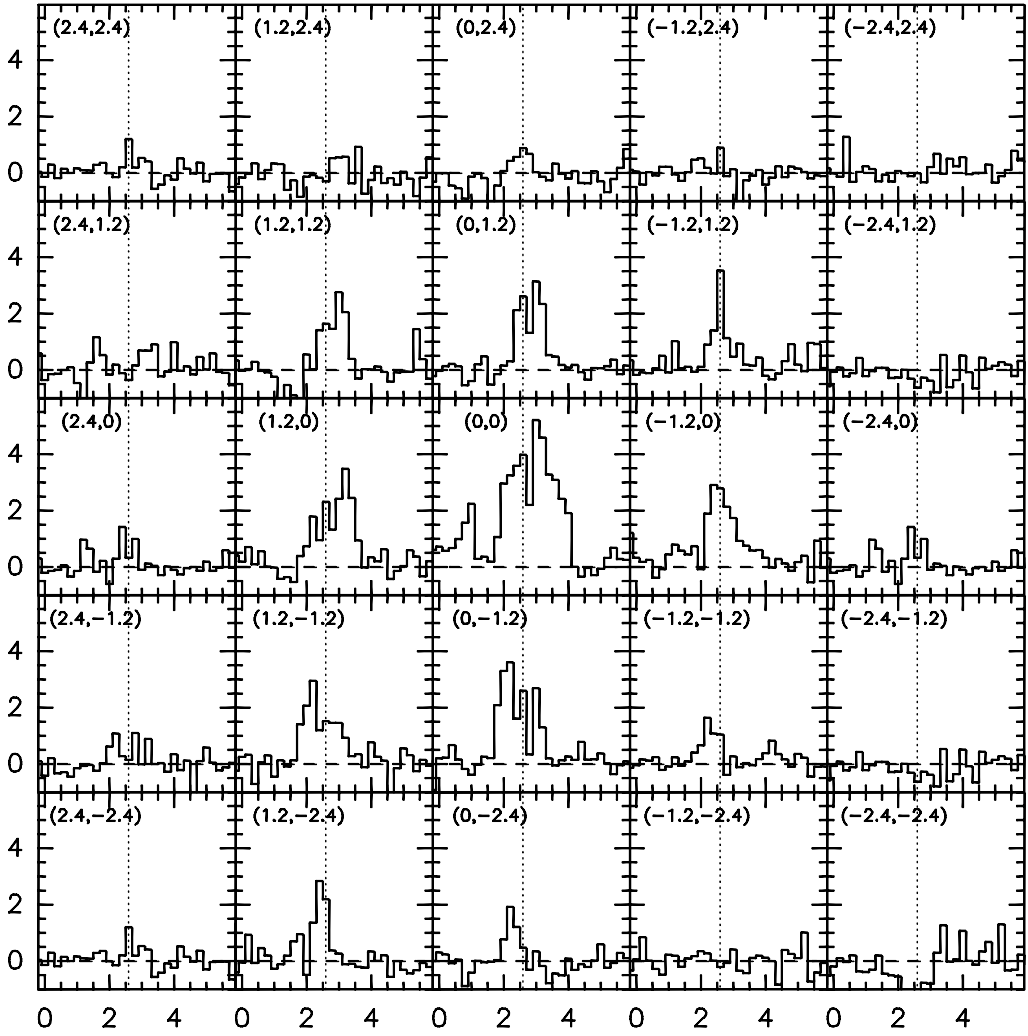}}
  \caption{ \label{fig7} C$^{18}$O~(\J{2}{1}) spectra obtained around
  the L1157-mm central region. The continuum has not been subtracted
  from the line emission. Offsets from the central position are
  indicated (in $''$) at the upper left corner of each plot. The
  dashed vertical line indicates the \vlsr\ of 2.6\kms.}
\end{figure}

Figure~\ref{fig6} presents the HC$_3$N~(\J{12}{11}) and
C$^{18}$O~(\J{2}{1}) spectra obtained at the position of L1157-mm. As
already mentioned, the C$^{18}$O emission shows a narrow ($\Delta
v\simeq0.2$\kms) deep absorption feature in the spectra, at a velocity
of $\sim$2.8\kms. A similar feature of $\sim$0.3\kms\ was also
reported at the same velocity in the interferometric
$^{13}$CO~(\J{1}{0}) map and spectra by Gueth et al.\ (1997). This
suggests that the C$^{18}$O~(\J{2}{1}) line is self-absorbed. However,
missing short-spacings information could cause a fake absorption, but
spatially extended (see Gueth et al.\ 1997, their Fig.7). In
Fig.~\ref{fig7} we plot the C$^{18}$O~(\J{2}{1}) spectra around
L1157-mm, by offsets of $1\farcs2$, roughly the size of our
synthesized beam. This figure shows that the absorption feature is
only detected at a few positions, close to the compact source and not
at larger distance of L1157-mm. Hence, the self-absorption, confined
in a small region ($\leq 5''$ or $\sim$2000~AU) around L1157-mm,
appears real. Note that Fig.~\ref{fig7} also shows line broadening
localized toward the center.

A redshifted self-absorption is naturally explained in a scenario
where an optically thick line is observed in an {\em infalling} core
with temperature decreasing outward (e.g.\ Leung \& Brown
1977). Fig.~\ref{fig6} shows that the brightness temperature (in
$T_{\rm mb}$ scale) of the self-absorbed feature is $\sim$2.2~K.
Assuming optically thick emission, it corresponds to an excitation
temperature of $T_{\rm ex}\simeq 6.3$~K, in excellent agreement with
the excitation temperature found by Gueth et al.\ (1997) from the
$^{13}$CO spectra. This low $T_{\rm ex}$ indicates that the
self-absorption features are most pro\-ba\-bly sub-thermally excited,
and should thus be produced in a medium whose density is low enough to
prevent thermalization of the lines. No self-absorption was detected
in the C$^{18}$O (\J{1}{0}) interferometric maps obtained by Gueth et
al.\ (1997). This suggests that the (\J{1}{0}) line is optically thin
while the (\J{2}{1}) is significantly thicker. For $T_{\rm ex}=10$~K,
the ratio of optical depths between the two transitions is
$\tau$(\J{2}{1})\,$\simeq1.9\,\tau$(\J{1}{0}). Note also that the
C$^{18}$O (\J{1}{0}) spectra from Gueth et al.\ (1997) has a narrow
profile, making it difficult to detect a possible asymetry.

\subsubsection{Other absorption features?}

In addition to the self-absorption feature discussed in the previous
paragraph, Fig.~\ref{fig6} reveals that both the HC$_3$N and C$^{18}$O
spectra at the position of L\,1157-mm have weaker emission in the
1--2\kms\ (blueshifted) and the 4--5\kms\ (redhifted) velocity
intervals. This suggests that the actual line profile may be
triangular-shaped, with two strong absorption features.  In both
lines, each of the ``absorbed'' channels is at a few $\sigma$ below
the un-affected emission, but the whole absorption feature is
significant because it extends over 3 to~6 adjacent channels. The
redshifted feature in the C$^{18}$O even shows an aborption of the
continuum emission.

\begin{figure}[!t]
\resizebox{\columnwidth}{!}{\includegraphics[angle=270]{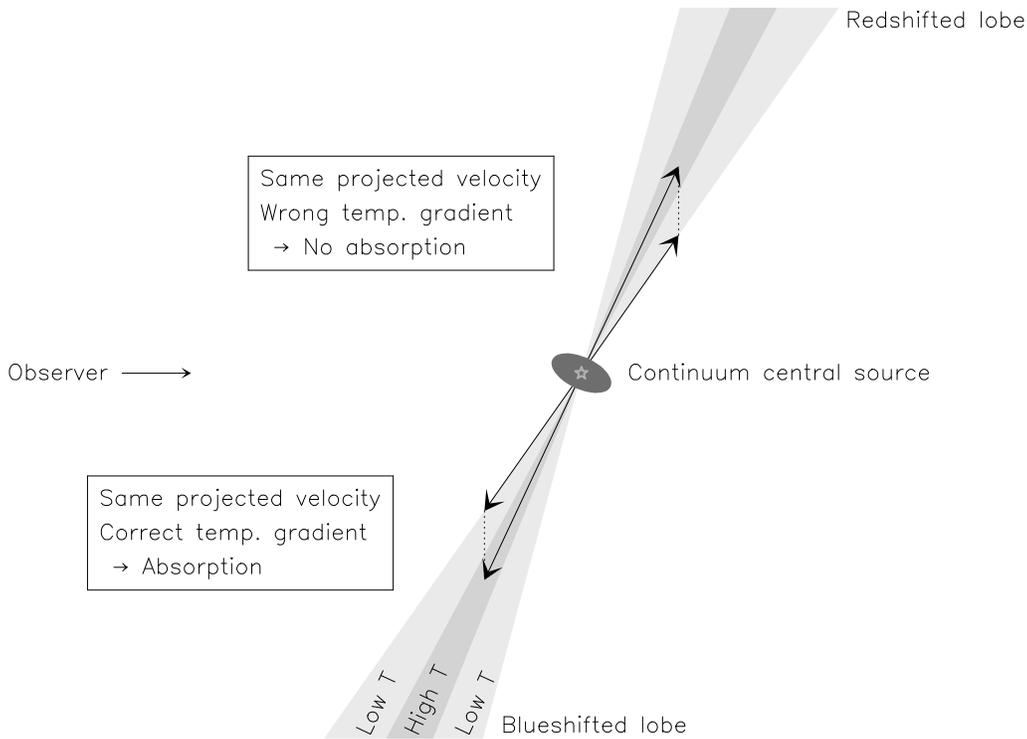}} 
  \caption{ \label{fig8} Geometry of the disk/outflow system in
	L\,1157 and possible origin of the {\em blueshifted}
	absorption feature.}
\end{figure}

To explain such absorptions in the spectra, one has to invoke the
presence of cold absorbing gas in front of a warm emitting
region. Concerning the {\em blueshifted} feature, it could arise from
self-absorption in the blueshifted southern outflow lobe (see
Fig.~\ref{fig8}): cold gas at the cavity edge is located in front of
warmer material present along the flow axis; the latter gas is
expected to have larger velocities, but since it is closer to the
plane of the sky, projection effects may produce the same
line-of-sight velocity for both components. As for the {\em
redshifted} absorption feature, a similar mechanism in the redshifted,
northern lobe cannot be proposed because the cold material would be
located behind the warm gas (Fig.~\ref{fig8}). Moreover, any
phenomenum in the redshifted outflow lobe could not explain the
absorption of the continuum, since this lobe is behind the central
source (which is at the origin of the compact continuum
emission). Instead, one could invoke absorption by the infalling
envelope, whose redshifted part is in front of the continuum source.

This latter scenario can drive interesting constrains on the size of
the continuum emitting region. The velocity at which the redshifted
continuum absorption is observed is higher than the self-absorption
feature: this means than the absorbing gas is located closer to the
center, and thus should be hotter. The excitation temperature should
thus be higher than 6.3~K. On another hand, the excitation temperature
must be significantly smaller than the continuum brightness
temperature (1~K), in order to create an absorption. These constrains
can be reconciled if we assume that the continuum emission is
dominated by a strongly peaked central component, which is affected by
beam dillution effects. Let's assume that the C$^{18}$O absorbing
material has an excitation temperature of 10~K. Judging from the
absorption feature (Fig.~\ref{fig6}), the continuum {\em true}
brightness temperature must be typically $\sim$5 times larger, i.e.\
$\sim$50~K. This can be explained by a beam dilution of $\sim$50 in
area, hence $\sim$7 in size. This in turn would imply a continuum
source size of $\sim$75~AU (beamsize/7$\simeq 0\farcs17$).
Interestingly, this value is of the order of the expected typical size
of a circumstellar disk in a Class~0 object (e.g.\ Pudritz et al.\
1996).

Clearly, these absorption features have to be confirmed and would
deserve further, more detailed, observations in order to derive a
clear understanding of the complex kinematics of the L1157
protostellar envelope.

\subsubsection{Rotation ?}

It should be mentioned that we did not find any evidence of velocity
gradient for the C$^{18}$O~(\J{2}{1}) transition (see
Fig.~\ref{fig4}), indicative of rotation, as suggested by Gueth et
al.\ (1997) for the C$^{18}$O~(\J{1}{0}) line.

\section{Summary}

We have studied with the IRAM Plateau de Bure millimeter
interferometer the dust and gas emission toward the core of the
Class~0 object L1157-mm. Our main conclusions are:

\begin{itemize}

\item The continuum emission at 2.7 and 1.3\,mm show a compact
component which is resolved at 1.3\,mm into a flattened compact core
of $\sim$450$\times250$~AU, and mass $\sim$0.12~$M_\sun$ and is
surrounded by an extended envelope of $\sim$3000~AU, and mass
$\sim$1.1~$M_\sun$.

\item The millimeter spectral index varies across the region. A lower
value $\alpha\simeq 2.1$ is found toward the position of the compact
protostar, while the spectral index is $\alpha\geq 4$ for the extended
surrounding material. These values imply a dust emissivity index
$\beta\simeq0.1$ and $\beta\geq 2$, respectively. This variation could
possibly indicate grain growth toward the compact component.

\item  A strong interaction between the molecular outflow and the
close protostellar environment is taking place and affects the
structure of the innermost parts of the envelope. This is shown by the
spatial correlation between the molecular outflow and the dust
continuum (1.3~mm map) and the HC$_3$N emission: both maps show
structures associated to the edges of the outflow lobes, as traced by
the $^{12}$CO emission.

\item HC$_3$N emission is also detected at a distance of
$\sim$13000~AU from the central object, at the position of a shock
associated to the impact of the precessing jet against the walls of
the main cavity of the southern lobe.

\item Evidence of infall has been detected through the
C$^{18}$O~(\J{2}{1}) observations, in agreement with the indications
of infall detected through $^{13}$CO by Gueth et al.\ (1997).

\item We did not detect any velocity gradient indicative of a rotating
circumstellar disk.

\end{itemize}

Finally, when observing Class~0 environment such as that of
L\,1157-mm, multiple velocity components and temperature gradients
along the line of sight do confuse the interpretation of spectral line
emission. In particular, they make it difficult to disentangle the
various components (disk, envelope, outflow) and determine their
velocity distributions. Our results show that even the highest density
gas tracers such as HC$_3$N are affected by the ejection phenemena,
tracing also ambient material compressed by the outflow. When deriving
results from low angular resolution observations of Class~0 objects
one should be aware of these problems. High angular resolution
observations of different dust and gas tracers are absolutely
necessary to get a detailed and accurate picture of such young objects
environment.

\begin{acknowledgements}
We would like to thank Paul Ho for his valuable comments on this
work. We acknowledge the IRAM staff from the Plateau de Bure for
carrying the observations. M.\ T.\ B.\ acknowledges support from a SAO
predoctoral fellowship.
\end{acknowledgements}

{}


\begin{thebibliography}{}
\bibitem[Adams, Lada, \& Shu (1987)]{Adams87} Adams, F.\ C., Lada, C.\ J., \& Shu, F.\ H.\ 1987, ApJ,
312, 788

\bibitem[Andr\'e, Ward-Thompson, \& Barsony (1993)]{andre93} Andr\'e, P., Ward-Thompson, D., \& Barsony, M.\ 1993, ApJ,
406, 122

\bibitem[Avery \& Chiao (1996)]{avery96} Avery, L.\ W., \& Chiao, M.\ 1996, ApJ, 463, 642

\bibitem[Bachiller et al.\ (1995)]{bachiller95} Bachiller, R., Liechti, S., Walmsley, C.\ M., \&
 Colomer, F.\ 1995, A\&A, 295, L51

\bibitem[Bachiller, Mart\'{\i}n-Pintado, \& Fuente (1993)]{bachiller93} Bachiller, R.,
Mart\'{\i}n-Pintado, J., \& Fuente, A.\ 1993, ApJ, 417, L45

\bibitem[Bachiller \& P\'erez Guti\'errez (1997)]{bachiller97} Bachiller, R., \& P\'erez Guti\'errez, M.\ 1997, ApJ, 487, L93

\bibitem[Bachiller et al.\ (2001)]{bachiller01} Bachiller, R., P\'erez Guti\'errez, M., Kumar, M.\ S.\ N., \& Tafalla, M.\
2001, A\&A, 372, 899

\bibitem[Beltr\'an et al.\ (2002)]{beltran02} Beltr\'an, M.\ T., Estalella, R., Ho, P.\ T.\ P., Calvet, N., Anglada, G.,
 Sep\'ulveda, I.\ 2002, ApJ, 565, 1069

\bibitem[Cassen \& Moosman (1981)]{cassen81} Cassen, P., \& Moosman, A.\ 1981, {\it Icarus}, 48, 353

\bibitem[Chini et al.\ (2001)]{chini01} Chini, R., Ward-Thompson, D., Kirk, J.\ M., Nielbock, M.,
Reipurth, B., \& Sie\-vers, A.\ 2001, A\&A, 369, 155

\bibitem[Chung, Osamu, \& Masaki (1991)]{chung91} Chung, H.\ S., Osamu, K., \& Masaki, M.\ 1991, JKAS, 24, 217

\bibitem[Davis \& Eisl\"offel (1995)]{davis95} Davis, C.\ J., \& Eisl\"offel, J.\ 1995, A\&A, 330, 851

\bibitem[Draine, \& Lee (1984)]{draine84} Draine, B.\ T., \& Lee, H.\ M.\ 1984, ApJ, 285, 89


\bibitem[Fuller et al.\ (1995)]{fuller95} Fuller, G.\ A., Ladd, E.\ F., Padman, R., Myers, P.\ C., Adams, F.\ C.\
1995, ApJ, 454, 862

\bibitem[Gueth, Guilloteau, \& Bachiller (1996)]{gueth96} Gueth, F., Guilloteau, S., \& Bachiller, R.\ 1996, A\&A, 307, 891

\bibitem[Gueth, Guilloteau, \& Bachiller (1998)]{gueth98} Gueth, F., Guilloteau, S., \& Bachiller, R.\ 1998, A\&A, 333, 287

\bibitem[Gueth et al.\ (1997)]{gueth97} Gueth, F., Guilloteau, S., Dutrey, A., \& Bachiller, R.\ 1997, A\&A, 323, 943

\bibitem[Gueth et al.\ (2003)]{gueth03} Gueth, F., Bachiller, R., \& Tafalla, M.\ 2003, A\&A, 401, L5

\bibitem[Guilloteau et al.\ (1992)]{guilloteau92} Guilloteau, S., Delannoy, J., Downes, D., et
al.\ 1992, A\&A, 262, 624

\bibitem[Henning, Michel, \& Stognienko (1995)]{henning95} Henning, Th., Michel, B., \& Stognienko, R.\ 1995, P\&SS, 43, 1333

\bibitem[Heyvaerts, J., \& Norman, C.\ (1989)]{heyvaerts89} Heyvaerts, J., \& Norman, C.\ 1989, ApJ, 347, 1055

\bibitem[Hirano, \& Taniguchi (2001)]{hirano01} Hirano, N., \& Taniguchi, Y.\ 2001, ApJ, 550, L219

\bibitem[Hodapp (1994)]{hodapp94} Hodapp, K.-W.\ 1994, ApJSS, 94, 615

\bibitem[Hogerheijde (2001)]{hogerheijde01} Hogerheijde, M.\ R.\ 2001, ApJ, 553, 618

\bibitem[Hogerheijde et al.\ (1998)]{hogerheijde98} Hogerheijde, M.\ R., van Dishoeck, E.\ F., Blake, G.\ A., \&
 van Langevelde, H.\ J.\ 1998, ApJ, 502, 315

\bibitem[Johnstone, \& Bally (1999)]{johnstone99} Johnstone, D., \& Bally, J.\ 1999, ApJ, 510, L49

\bibitem[K\"onigl \& Pudritz  (2000)]{konigl00} K\"onigl, A., \& Pudritz R.\ E.\ 2000, in {\it
Protostars and Planets IV}, eds.\ V.\ Mannings, A,\ P.\ Boss, \& S.\ S.\ Russell, (Tucson:
University of Arizona Press), 759

\bibitem[Ladd et al.\ (1995)]{ladd95} Ladd, E.\ F., Fuller, G.\ A., Padman, R., Myers, P.\
C., \& Adams, F.\ C.\ 1995, ApJ, 439, 771

\bibitem[Lai \& Crutcher (2000)]{lai00} Lai, S.-P., \& Crutcher, R.\ M.\ 2000, ApJSS, 128, 271

\bibitem[Langer, Velusamy, \& Xie (1996)]{langer96} Langer, W.\ D., Velusamy, T., \& Xie, T.\ 1996, ApJ, 468, L41

\bibitem[Larson (1969)]{larson69} Larson, R.\ B.\ 1969, MNRAS, 145, 271

\bibitem[Leung \& Brown (1977)]{leung77} Leung, C., \& Brown, R.\ 1977, ApJ, 214, L73

\bibitem[Mannings \& Emerson (1994)]{mannings94} Mannings, V., \& Emerson, J.\ P.\ 1994, MNRAS, 267, 361

\bibitem[Motte \& Andr\'e (2001)]{motte01} Motte, F., \& Andr\'e, P. 2001, A\&A, 365, 440

\bibitem[Ohashi et el.\ (1997)]{ohashi97} Ohashi, N., Hayashi, M., Ho, P.\ T.\ P., \& Momose,
M.\ 1997, ApJ, 475, 211

\bibitem[Ohashi et al.\ (1999)]{ohashi99} Ohashi, N., Lee, S.\ W., Wilner, D.\ J., \& Hayashi,
M.\ 1999, ApJ, 518, L41

\bibitem[Ossenkopf \& Henning (1994)]{ossenkopf94} Ossenkopf, V., \& Henning, T. 1994, A\&A, 291, 943

\bibitem[Pollack et al.\ (1994)]{pollack94} Pollack, J.\ B., Hollenback, D., Beckwith, S.\ W.\ V., Simonelli, D.\ P.,
Roush, T., \& Fong, W. 1994, ApJ, 421, 615

\bibitem[Pudritz \& Norman (1983)]{pudritz83} Pudritz, R.\ E., \& Norman, C.\ A.\ 1983, ApJ, 274, 677

\bibitem[Pudritz et al.\ (1996)]{pudritz96} Pudritz, R.\ E., Wilson, C.\ D., Carlstrom J.\ E, et al., 1996, ApJ, 470, L123

\bibitem[Saito et al.\ (1996)]{saito96} Saito, M., Kawabe, R., Kitamura, Y., \& Sunada, K.\
1996, ApJ, 473, 464

\bibitem[Scoville et al.\ (1986)]{scoville86} Scoville, N.\ Z., Sargent, A.\ I., Sanders, D.\ B., Claussen,
M.\ J., Masson, C.\ R., Lo, K.\ Y., \& Phillips, T.\ G.\ 1986, ApJ, 303, 416

\bibitem[Shirley et al.\ (2000)]{shirley00} Shirley, Y.\ L., Evans, N.\ J., Rawlings, J.\ M.\ C., \&
Gregersen, E.\ M. 2000, ApJSS, 131, 249

\bibitem[Shu, Adams, \& Lizano (1987)]{shu87} Shu, F.\ H., Adams, F.\ C.\, \& Lizano, S.\ 1987, ARAA, 25,
23

\bibitem[Shu et al.\ (1994)]{shu94} Shu, F., Najita, J., Ostriker, E., Wilkin, F., Ruden, S.,
\& Lizano, S.\ 1994, ApJ, 429, 781

\bibitem[Smith et al.\ (2000)]{smith00} Smith, K.\ W., Bonnell, I.\ A., Emerson, J.\ P., \& Jenness,
 T.\ 2000, MNRAS, 319, 991

\bibitem[Tafalla \& Bachiller (1995)]{tafalla95} Tafalla, M., \& Bachiller, R.\
1995, ApJ, 443, L37

\bibitem[Terebey, Shu, \& Cassen (1984)]{terebey84} Terebey, S., Shu, F.\ H., \& Cassen, P.\ 1984, ApJ, 286, 529

\bibitem[Umemoto et al.\ (1992)]{umemoto92} Umemoto, T., Iwata, T., Fukui, Y., Mikami, H., Yamamoto, S., Kameya,
 O., \& Hirano, N.\ 1992, ApJ, 392, L83

\bibitem[Visser et al.\ (1998)]{visser98} Visser, A.\ E., Richer, J.\ S., Chandler, C.\ J., \& Padman,
R.\ 1998, MNRAS, 301, 585

\bibitem[Zhang, Ho, \& Wright (2000)]{zhang00} Zhang, Q., Ho, P.\ T.\ P., \& Wright, M.\ C.\ H.\ 2000, AJ, 119, 1345

\bibitem[Zhang et al.\ (1995)]{zhang95} Zhang, Q., Ho, P.\ T.\ P., Wright, M.\ C.\ H., \&
 Wilner, D.\ J.\ 1995, ApJ, 451, L71

\end{thebibliography}
\end{document}